\documentclass[aps,prl,twocolumn,showpacs,amssymb,floatfix,superscriptaddress]{revtex4-1}
\usepackage[utf8]{inputenc}
\usepackage[english]{babel}

\usepackage{hyperref}
\usepackage{color}
\usepackage[usenames,dvipsnames]{xcolor}
\usepackage{verbatim}

\usepackage{amsmath}
\usepackage{amsfonts}
\usepackage{amssymb}
\usepackage{graphicx}
\usepackage{lmodern}
\usepackage{booktabs}
\usepackage{tikz}	
\usetikzlibrary{arrows}

\usepackage{url}

\usepackage[left=2cm,right=2cm,top=2cm,bottom=2.2cm]{geometry}

\usepackage{hyperref}
\usepackage{color}
\usepackage[usenames,dvipsnames]{xcolor}
\hypersetup{colorlinks=true, citecolor=orange, urlcolor=cyan, linkcolor=blue}

\usepackage{titlesec} 

\begin{document}

\title{\large{\textbf{Entropy-based randomisation of rating networks}}}
\author{Carolina Becatti}
\affiliation{IMT School for Advanced Studies, Piazza S.Francesco 19, 55100 Lucca - Italy}
\author{Guido Caldarelli}
\affiliation{IMT School for Advanced Studies, Piazza S.Francesco 19, 55100 Lucca - Italy}
\affiliation{Istituto dei Sistemi Complessi (ISC)-CNR UoS Universit\`a ``Sapienza'', Piazzale Aldo Moro 5, 00185 Rome - Italy}
\affiliation{ECLT San Marco 2940, 30124 Venezia , Italy}
\author{Fabio Saracco}
\affiliation{IMT School for Advanced Studies, Piazza S.Francesco 19, 55100 Lucca - Italy}

\begin{abstract}
In the last years, due to the great diffusion of e-commerce, online rating platforms quickly became a
common tool for purchase recommendations. However, instruments for their analysis did not evolve
at the same speed. Indeed, interesting information about users’ habits and tastes can be recovered
just considering the bipartite network of users and products, in which links have different weights
due to the score assigned to items. With respect to other weighted bipartite networks, in these
systems we observe a maximum possible weight per link, that limits the variability of the outcomes.
In the present article we propose an entropy-based randomisation of (bipartite) rating networks by
extending the Configuration Model framework: the randomised network satisfies the constraints of
the degree per rating, i.e. the number of given ratings received by the specified product or assigned
by the single user. We first show that such a null model is able to reproduce several non-trivial
features of the real network better than other null models. Then, using it as a benchmark, we
project the information contained in the real system on one of the layers, showing, for instance,
the division in communities of music albums due to the taste of customers, or, in movies due the
audience.
\end{abstract}

\maketitle

\section{Introduction}
Network theory \cite{caldarelli2007scale-free,newman2010networks} proved successful \cite{barabasi2009scale-free} in the description and modelling of a wide variety of systems, ranging from the obvious cases of the Internet \cite{pastor2001dynamical,vazquez2001large}, the  WWW \cite{meusel2015graph} and social networks \cite{gonzalez2011dynamics}. In these settings they formed the evidence on which computational social science is based \cite{lazer2009social}, to cell properties in biology \cite{buchanan2010networks} and fMRI imaging in brain analysis \cite{avena2017communication,mastrandrea2017organization} contributing to the new field of network medicine \cite{scala2012using,greene2017putting}, to banks in financial systems \cite{demasi2006fitness,vodenska2017systemic}. Networks come in various shapes, from the simplest case of similar vertices connected by binary edges, to weighted and/or directed networks, to multigraphs where more than one edge can connect two vertices, to bipartite graphs where two distinct sets of vertices are present. Simple examples of the latter case are bipartite graphs in which a connection is drawn if an individual (on one set) performs or not a given task (in the other set). Here we focus on a specific case of rating networks, where the sets are those of individuals and products, while the edges represent reviews of products by consumers and are weighted by the numerical score received (as for example the well-known Amazon review system). 

As in the case of ordinary networks the question is the assessment of the significance of the topological quantities measured. This means that in order to consider ``relevant'' any particular value measured, it should ``substantially" deviate from a ``random'' realization of the same network. The problem of course is how to define what ``random" means in this case, and then how ``substantial" a deviation is.
Following a nowadays relevant stream of literature \cite{park2004statistical,garlaschelli2008maximum,squartini2011analytical}, we answer to the former question by  defining  an appropriate ensemble of graphs from which we can obtain a benchmark. 
This procedure reveals to be an extremely powerful instrument for the analysis of many non-trivial network properties. In a nutshell, the method prescribes to define a probability distribution over the ensemble, through a constrained entropy maximization procedure. Then, the maximization of the related likelihood function provides the probability that any possible pair of nodes in the network of interest is connected. The constraints introduced in the first maximization procedure are the topological quantities of the real network, i.e. for binary and undirected networks the degree of each node is used as a constraint. Once we have a theoretical framework, we can even state, by comparing the actual observations with the expectation of the null model, if the real values deviate substantially from the theoretical distribution.

A more restrictive null model permit to reproduce with a higher accuracy the feature of the original network, thus capturing more the features of the real network. In this paper we constrain not only the presence of positive reviews, but even the exact ratings; due to its application, we indicate it in the following as Bipartite Score Configuration Model, \emph{BiSCM}. 

Rating networks may be interpreted as classical weighted networks, whose edges are weighted by a finite set of discrete scores. In this context, appropriate constraints are represented by the specification of nodes' strengths only (Weighted Configuration Model, WCM,  \cite{squartini2011analytical}). Because of the extremely poor predictive power of vertices' strengths, an ``enhanced'' version of the previous model has also been introduced (Enhanced Configuration Model, ECM) in \cite{mastrandrea2014reconstructing}; this method adds the topology as additional information. 
The presence of a finite number of discrete weights 
complicates the problem formulation and extremely 
the required computational effort. For these reasons, a preliminar ``binarization'' procedure is often employed (it is the approach of \cite{saracco2017inferring}, but it is usual even in recommendation systems, like in, for instance, \cite{Rennie2005}), by thresholding the weight of the edges. In this way, the resulting network is binary and can be easily randomized with the Bipartite Configuration Model (BiCM) in \cite{saracco2015randomizing}. The peculiarity of our method is that we avoid the scores-related problems by specifying a \emph{multi-degree} for each node in the network, i.e. by ~\emph{specifying the entire distribution of scores received by a node}. 

The rest of this paper is structured as follows. In the Method section we explain in details the entire ensemble construction procedure. In the Datasets section we briefly review the datasets used to test the method. The main results are reported in the Results section, where we describe all the performed analyses. We finally discuss any advantages and drawbacks of the method in the Discussion section.

\section{Methods}
\begin{figure}
\centerline{
\includegraphics[width=8cm]{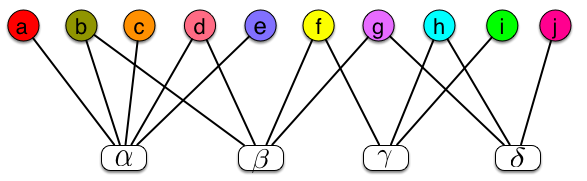}}
\caption{A simple bipartite graph. In the following Latin letters will indicate users, and Greek letters goods.}
\label{fig:bip}
\end{figure}

A bipartite rating network is a network that can be partitioned in two sets such that the edges of the network are only between a vertex of the first set with a vertex of the second set (see Figure \ref{fig:bip}). 
This kind of structure arises naturally whenever considering collaboration networks (i.e. actors in the first set, movies on the second set), export of products, consumers and goods etc. To distinguish the two sets, the index running on one set ($L$) is typically indicated by a Latin letter $i,j,k$ and the index running on the second set ($\Lambda$) is indicated by a Greek letter $\alpha, \beta, 
\gamma ...$.    A bipartite network with $N = N_L + N_\Lambda$ vertices and $E$ edges can be entirely specified by its $N_L \times N_\Lambda$ adjacency matrix \textbf{M} with entries $m_{i,\alpha} = \beta$, whenever product $\alpha$ has been reviewed and assigned score $\beta$ by user $i$, and $m_{i,\alpha} = 0$ otherwise. In what follows, we only deal with the case in which users are required to assign discrete scores and the maximum possible score is known, denoted from now on as $\beta_{max}$. All members of the benchmark ensemble will have a constant number of vertices for each layer, respectively equal to $N_L$ and $N_\Lambda$. A binary representation of \textbf{M}'s entries can be considered, defining $m_{i,\alpha,\beta} = \delta(m_{i,\alpha}, \beta)$ for $\beta = 1, \dots, \beta_{max}$, where $\delta$ is the classical Kronecker delta function. By doing so, the variable $m_{i,\alpha,\beta}$ will be equal to $1$ if node $i$ has reviewed node $\alpha$ with the numerical score $\beta$ and $m_{i,\alpha,\beta} = 0$ otherwise.

We use the notation 
\begin{equation}
k_{i,\beta}(\textbf{M}) = \sum_\alpha m_{i,\alpha,\beta} \;\;\; i=1, \dots, N_L
\label{eq:degree-a}
\end{equation}
\begin{equation}
\label{eq:degrees}
k_{\alpha,\beta}(\textbf{M}) = \sum_im_{i,\alpha,\beta} \;\;\; \alpha = 1, \dots, N_\Lambda
\end{equation}
to indicate the number of reviews with score $\beta$ respectively assigned by a generic user $i$ (eq. (\ref{eq:degree-a})) and received by a generic product $\alpha$ (eq. (\ref{eq:degrees})) respectively. The specification of eqs. (\ref{eq:degree-a}-\ref{eq:degrees}) for all $\beta = 1, \dots, \beta_{max}$ defines the distribution of scores received by each node and constitute the fundamental constraints of our problem. Note that this framework can also be intended as describing a multiedge network in which $\beta_\text{max}$ is the maximum number of edges allowed between any couple of nodes.

At this point we look for the instances of the graph maximising the (Shannon's) Entropy 
\begin{equation}\label{eq:entropy}
S = - \sum_\textbf{M} P(\textbf{M}) \ln P(\textbf{M})
\end{equation}
under the constraints $\left\langle k_{i,\beta} \right\rangle = k_{i,\beta}$ and $\left\langle k_{\alpha,\beta} \right\rangle = k_{\alpha,\beta}$ for $i = 1, \dots, N_L$, $\alpha = 1, \dots, N_\Lambda$ with $\beta = 1, \dots, \beta_{max}$. In other words, we consider the probability distribution over the ensemble, such that the expected degree of each node, for every possible ratings, equals, on average, its observed value, while keeping all the rest maximally random.

The solution to this bipartite maximization problem gives the following probability distribution over the ensemble
\begin{equation}\label{eq:P_M}
P(\mathbf{M}|\vec{x},\vec{y}) = \prod_{i,\alpha} q_{i,\alpha}(m_{i,\alpha,\beta}|\vec{x},\vec{y}) 
\end{equation}
where $\vec{x}$ is a $N_L\beta_{max}$ vector of Lagrange multipliers that controls the expected degrees for each possible rating for users, while $\vec{y}$ is the analogous $N_\Lambda\beta_{max}$ dimensioned vector of Lagrangian multipliers for the products 
and
\begin{equation}\label{eq:q_m_ia}
q_{i,\alpha}(m_{i,\alpha,\beta}|\vec{x},\vec{y}) = \frac{\prod_\beta (x_{i,\beta}\, y_{\alpha,\beta})^{m_{i,\alpha,\beta}}}{1+\sum_\beta x_{i,\beta}\, y_{\alpha,\beta}}
\end{equation}
is the probability to observe a link between nodes $i$ and $\alpha$ (see Appendix for further details). 
At this point for every node we can assign a vectorial Lagrangian multiplier ($\vec{x}_{i}$ if it belongs to the layer L, $\vec{y}_{\alpha}$ if it belongs to the layer $\Lambda$) of dimension $\beta_{max}$. Thus the probability to observe a link with rating $\beta$ can be expressed as, as
\begin{equation}\label{eq:probs}
p_{i,\alpha,\beta} = \frac{x_{i,\beta}\, y_{\alpha,\beta}}{1 + \sum_\beta x_{i,\beta}\, y_{\alpha,\beta}}
\end{equation}
for all $i$, $\alpha$ and $\beta$. 


In order to determine the numerical values for our Lagrange multipliers, let us consider a specific real-world rating network $\mathbf{M}^*$, for which the degree sequence $\{k_{i,\beta}(\mathbf{M}^*), k_{\alpha,\beta}(\mathbf{M}^*)\}$ is known for all $i, \alpha$ and for each rating $\beta$. The log-likelihood defined by equation~(\ref{eq:P_M}) is given by
\begin{eqnarray}\label{eq:likelihood}
\mathcal{L}(\vec{x},\vec{y}|\mathbf{M}^*) &=& 
\sum_{i,\beta} k_{i,\beta}(\textbf{M}^*) \ln x_{i,\beta} \nonumber \\
&+& \sum_{\alpha,\beta} k_{\alpha,\beta}(\textbf{M}^*) \ln y_{\alpha,\beta} \nonumber \\
&-&\sum_{i,\alpha}\ln \Big(1 + \sum_\beta x_{i,\beta}\,y_{\alpha,\beta}\Big)
\end{eqnarray}
The maximization procedure consists then in finding the specific parameter values $\vec{x}$ and $\vec{y}$ that maximizes the probability to observe the network of interest $\textbf{M}^*$. 
Thus, the benchmark model for the real-world network $\mathbf{M}^*$ is completely specified and it is possible to compare its observed topological properties with the same quantities averaged over the ensemble of graphs. 

Let us conclude this section with some remarks: the null model's calibration (i.e.~the determination of the Lagrange multipliers vector $\vec{x}$ and $\vec{y}$) may easily become costly, since the number of equations and unknowns of the system 
grows linearly with the number of nodes in the network $N$ and observed scores $\beta_{max}$. For this reason, for extremely large and sparse systems an approximation is provided in the Appendix. As in the  Chung-Lu model \cite{chung2002connected}, we relax the constraints required by considering expected values and we approximate the Lagrangian multipliers to be proportional to the degree of the node for the given score. We will show that for high degrees, this approximation systematically overestimates the exact probability, but there is a quite good agreement for low degrees nodes.
\newline Consider that our framework permits to randomise also categorical data and signed networks. Indeed, so far there is no hypothesis about the nature of the different outputs of the adjacency matrix $\textbf{M}$, but for the fact that they are mutually exclusive. In the case of signed networks, different $\beta$s would have been $0$ (absence of any kind of link), $+1$ (presence of a positive link) and $-1$ (presence of a negative link) and the matrices $\textbf{M}^\beta$ would have been defined therefore. In the case of mutually exclusive categories, the story is the same, just assigning to each $\beta$ a different category.

\subsection{Higher order topological benchmark}
In the cases of study considered we can distinguish ``positive'' from ``negative'' reviews. In the ML network described below in Section Datasets, the count of negative reviews is large enough to make a proper analysis of this information valuable. Therefore, after the randomization, we can define a signed version of the original adjacency matrix. This new matrix $\overline{\mathbf{M}}$, has entries $m^+_{i,\alpha}$ or $m^-_{i,\alpha}$ (i.e. $\overline{m}_{i,\alpha} = +1$ or $\overline{m}_{i,\alpha} = -1$) whenever a positive or, respectively, negative review was registered in $\mathbf{M}^*$. We denote the quantities $k_i^+ = \sum_\alpha m_{i,\alpha}^+$ and $k_i^- = \sum_\alpha m_{i,\alpha}^-$ respectively \textit{positive degree} and \textit{negative degree}, to indicate the number of edges with positive or negative sign incident to node $i$. The previous quantities are equivalently defined for node $\alpha$. Finally, the related probability matrices have been indicated as $\left\langle\textbf{M}^+\right\rangle$ and  $\left\langle\textbf{M}^-\right\rangle$, whose entries $\left\langle m_{i,\alpha}^+\right\rangle$ and $\left\langle m_{i,\alpha}^-\right\rangle$ represent the probabilities that user $i$ positively or negatively reviews node $\alpha$ (i.e.~with scores $\beta = 3, 4, 5$ or $\beta = 1, 2$). 

On the dataset, we first analyze the correlation between neighbor nodes' degrees, introducing a signed version of the classical average nearest neighbor degree (ANND). We separately analyze all possible combinations of positive/negative neighbors and positive/negative degrees, as follows
\begin{eqnarray}
\label{eq:annd_pp}
k_i^{pp}(\overline{\textbf{M}}) &=& \frac{\sum_\alpha m^+_{i,\alpha} k_\alpha^+}{k_i^+} \nonumber \\
\label{eq:annd_pn}
k_i^{pn}(\overline{\textbf{M}}) &=& \frac{\sum_\alpha m^+_{i,\alpha} k_\alpha^-}{k_i^-} \nonumber \\
\label{eq:annd_np}
k_i^{np}(\overline{\textbf{M}}) &=& \frac{\sum_\alpha  m^-_{i,\alpha} k_\alpha^+}{k_i^+} \nonumber \\
\label{eq:annd_nn}
k_i^{nn}(\overline{\textbf{M}}) &=& \frac{\sum_\alpha m^-_{i,\alpha} k_\alpha^-}{k_i^-}.  
\end{eqnarray}

\begin{figure}
\centering
\begin{tikzpicture}[>=stealth',shorten >=1pt,auto,node distance=2cm,
                    thick,main node/.style={circle,draw,font=\sffamily\small\bfseries}]
  \node[main node] (1) {$i$};
  \node[main node] (2) [right of=1] {$j$};
  \node[main node] (3) [below of=1] {$\alpha$};
  \node[main node] (4) [right of=3] {$\beta$};
  
    \path[every node/.style={font=\sffamily\small}]
    (1) edge node {} (4)
    (2) edge node {} (3);
    \path[every node/.style={font=\sffamily\small}, dashed]
    (1) edge node {} (3)
    (2) edge node {} (4);
\end{tikzpicture}
\qquad
\begin{tikzpicture}[>=stealth',shorten >=1pt,auto,node distance=2cm,
                    thick,main node/.style={circle,draw,font=\sffamily\small\bfseries}]
  \node[main node] (1) {$i$};
  \node[main node] (2) [right of=1] {$j$};
  \node[main node] (3) [below of=1] {$\beta$};
  \node[main node] (4) [right of=3] {$\alpha$};
  
    \path[every node/.style={font=\sffamily\small}]
    (1) edge node {} (3)
    (2) edge node {} (4);
    \path[every node/.style={font=\sffamily\small}, dashed]
    (1) edge node {} (4)
    (2) edge node {} (3);
\end{tikzpicture}
\caption{Checkerboard-like motifs.}
\label{fig:tikz}
\end{figure}
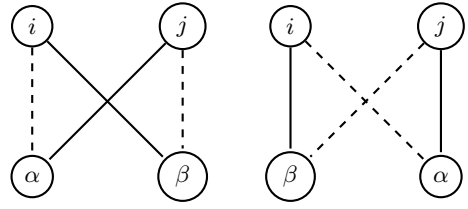

The first apex letter is referred to the sign of the edges incident to $i$, while the second one indicates the sign of $i$'s neighbor degree. Then, we compute the number of signed checkerboard-like motifs $c_i$ (Ref) for each node, as follows
\begin{eqnarray}
c_i(\overline{\textbf{M}}) &=& \sum_\alpha \sum_{\beta} \sum_{j} m_{i,\beta}^+ m_{j,\alpha}^+ m_{i,\alpha}^-m_{j,\beta}^- 
\label{eq:checkerboard}
\end{eqnarray}
Equations~(\ref{eq:annd_pp}-\ref{eq:checkerboard}) are analogously defined for column nodes $\alpha$. The number of checkerboard-like motifs node $i$ is involved represents the number of times user $i$ disagrees with another user $j$ about the reviews assigned to a pair of movies $(\alpha, \beta)$. Their graphical representation is provided in Figure \ref{fig:tikz}, where continuous edges represent positive reviews while dashed represent negative ones.

\section{\textbf{Monopartite projection}}
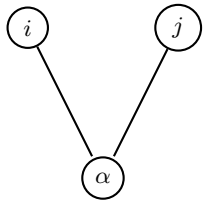
\begin{figure}[hb!]
\centering
\begin{tikzpicture}[>=stealth',shorten >=1pt,auto,node distance=2cm,
                    thick,main node/.style={circle,draw,font=\sffamily\small\bfseries}]
  \node[main node] (1) at (0,2) {$i$} ;
  \node[main node] (3) at (1,0) {$\alpha$} ;
  \node[main node] (2) at (2,2) {$j$} ;

    \path[every node/.style={font=\sffamily\small}]
    (1) edge node {} (3)
    (2) edge node {} (3);
\end{tikzpicture}
\caption{A V-motif.}
\label{fig:vmotif}
\end{figure}

The traditional way to analyse collaboration systems \cite{barabasi1999emergence} (e.g. actors in the movie system), it is to project the information contained in a bipartite network on one of the layers by considering the statistical significance of their common connections. From various attempts on boards \cite{caldarelli2004corporate} to more recent approaches \cite{tumminello2011statistically}, several works \cite{gualdi2016statistically,dianati2016maximum,saracco2017inferring,straka2017grand} applied a similar idea, making now use of the (bipartite) Configuration Model. Summarising, once the probability for the single bipartite link is calculated, it is possible to compute the probability that a pair of nodes shares a link with an item on the opposite layer. Such a pattern can be represented by a V-motif (see Figure~\ref{fig:vmotif}): if the probabilities per link are independent, the probability of observing the single V-motif of Figure~\ref{fig:vmotif} is simply $P(V^{i,j}_\alpha)=p_{i,\alpha}p_{j,\alpha}$. Thus, the probability to observe a certain number of V-motifs between $i$ and $j$, follows the Poisson (Binomial) distribution \cite{volkova1996refinement,hong2013computing}, i.e. the distribution of $N_\Lambda$ independent Bernoulli events, each with different, in general, probabilities. Comparing the observation on the real network with their theoretical Poisson Binomial distribution, it is possible to calculate a p-value for each pair of nodes on the same layer. After a multiple hypothesis testing procedure it is possible to state which connections of the monopartite projection are statistically relevant, i.e. which are the nodes that share more connections than expected by the null model.

\begin{table*}[ht!]
\centering
\begin{tabular}{lcccccccc}
\hline \hline
 & $N$ & $N_L$ & $N_\Gamma$ & $E$ & $\rho$ & $+$ & $-$\\
\hline
(ML) MovieLens & 2,588 & 1,645 & 943 & 100,000 & $6.36 \cdot 10^{-2}$ & 0.83 & 0.17 \\
(MI)  Musical Instruments & 2,329 & 900 & 1,429 & 10,261 & $7.98 \cdot 10^{-3}$ & 0.95 & 0.05 \\
(SM)  Smartphones & 16,172 & 2,256 & 13,916 & 15,817 & $5.04 \cdot 10^{-4}$ & 0.73 & 0.27\\
(DM) Digital Music & 7,000 & 2,500 & 4,500 & 36,774 & $3.27 \cdot 10^{-3}$ & 0.91 & 0.09\\
\hline \hline
\end{tabular}
\caption{Data description}
\label{table:table_data}
\end{table*}

In the present paper, we shall extend such a procedure to rating networks, considering couples of items that receive both positive reviews from the same customer. The probability of having a positive review, if we set the threshold for positive review at $\beta^*$, is given by 
\begin{equation*}
p_{i,\alpha}^+=\sum_{\beta=\beta^*}^{\beta_{max}}p_{i,\alpha, \beta}.
\end{equation*}
Then, the algorithm follows exactly the same steps of the original procedure: we consider the probability that the same users give a positive review to both of the items and calculate the distribution of common good reviews  as above. Although several methods are available in the literature, we employ the ``false discovery rate'' procedure \cite{benjamini1995controlling} to validate the previously calculated p-values, since it permits to have a stricter control on the false positives. The result of the algorithm is a threshold p-value, used to validate all the hypotheses in matrix (\ref{eq:z}) at a time. For a quick recap, the method's recipe is the following: 
\begin{itemize}
\item sort the vector of p-values to be tested in ascending order, 
\item select the largest integer $\hat{i}$ satisfying
\begin{equation}\label{eq:cond_fdr}
\hbox{p-value}_{\hat{i}} \leq \frac{\hat{i}t}{N_L N_\Lambda},
\end{equation}
where $t$ is the chosen significance level, 
\item consider $\hbox{p-value}_{\hat{i}}$ as the threshold value. 
\end{itemize}
Then, all hypotheses whose p-value is smaller than or equal to the threshold must be rejected, while we are not able to reject all hypotheses whose p-value is greater than $\hbox{p-value}_{\hat{i}}$. In the whole paper we will consider $t=0.05$.

\section{Datasets}
The following datasets have been employed to test the method. 
\label{datasets}
\begin{itemize}
\item \textbf{MovieLens 100k}: Bipartite network that collects 100,000 movies' ratings. The website's users are characterized by some individual features, such as age, job, sex, state and zipcode. For the set of movies we have information on the release year, title and genre. Each user can review a movie with a score that ranges from $1$ to $5$, according to her level of appreciation. The data has been downloaded from the repository \footnote{\url{http://konect.uni-koblenz.de/networks/}}\setcounter{footnote}{1}, while any additional information is provided in \cite{resnick1994open}.
\item \textbf{Amazon}: We collected three datasets involving different cathegories of products. From \footnote{\url{http://jmcauley.ucsd.edu/data/amazon/}} we downloaded (from the \textit{``Small'' subsets for experimentation} section) the Musical Instruments and Digital Music datasets, that respectively collect purchases of musical instruments and CDs or vinyls (the latter had to be further sampled due to its high dimensions). The data about Smartphones and related products has instead been downloaded from \footnote{\url{http://times.cs.uiuc.edu/~wang296/Data/}}. For all of them, the possible ratings for each purchased product span from $1$ to $ 5$.
\end{itemize}
A more detailed description of the datasets is provided in Table \ref{table:table_data}, where $\rho$ denotes the density of the observed network while the symbols $+$ and $-$ indicates the percentage of positive and negative edges in each dataset.

\section{Results}

\begin{figure*}[ht!]
\center
\includegraphics[width = 3.9cm, height=3.9cm]{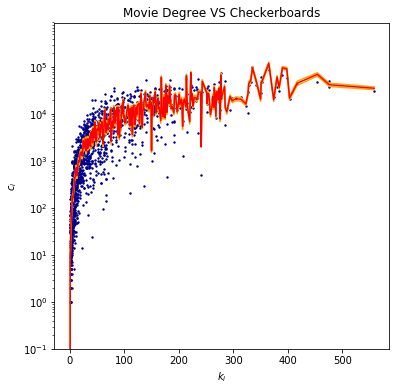}
\includegraphics[width = 3.9cm, height=3.9cm]{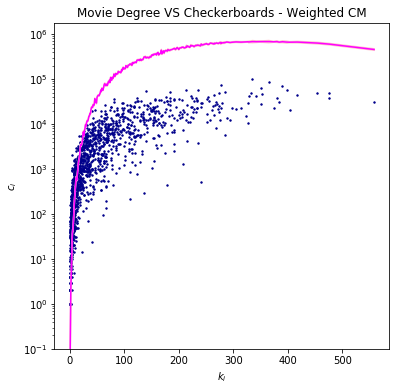}
\includegraphics[width = 3.9cm, height=3.9cm]{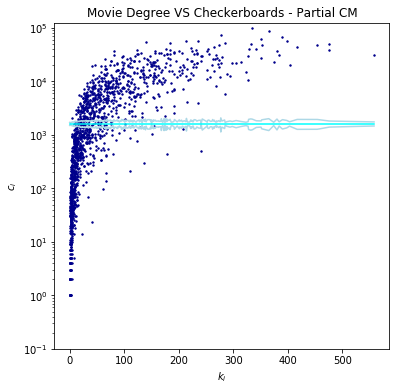}
\includegraphics[width = 3.9cm, height=3.9cm]{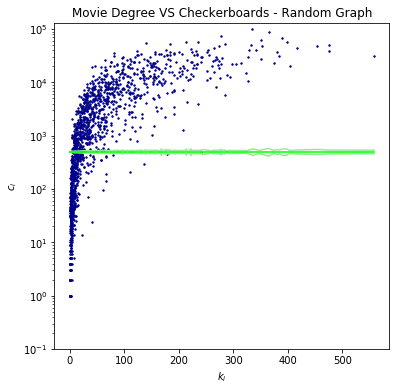}\\
\includegraphics[width = 3.9cm, height=3.9cm]{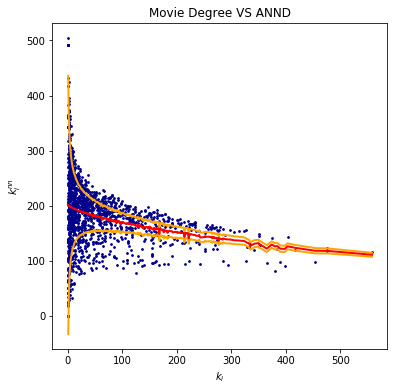}
\includegraphics[width = 3.9cm, height=3.9cm]{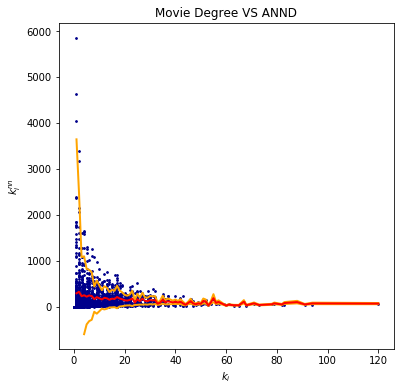}
\includegraphics[width = 3.9cm, height=3.9cm]{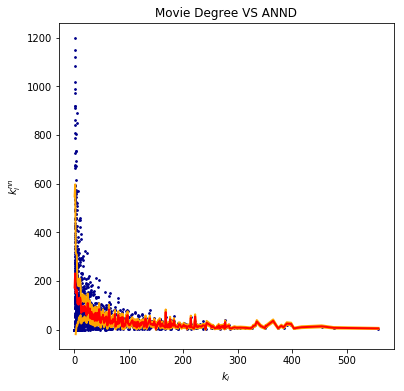}
\includegraphics[width = 3.9cm, height=3.9cm]{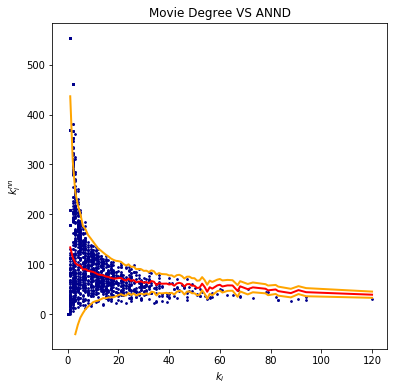}\\
\includegraphics[width = 3.9cm, height=3.9cm]{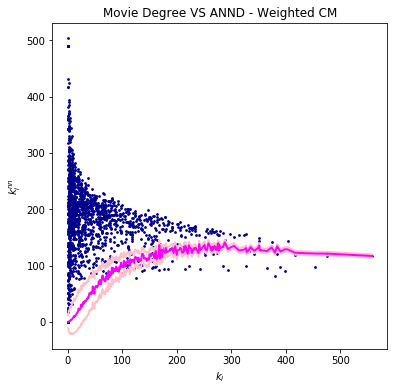}
\includegraphics[width = 3.9cm, height=3.9cm]{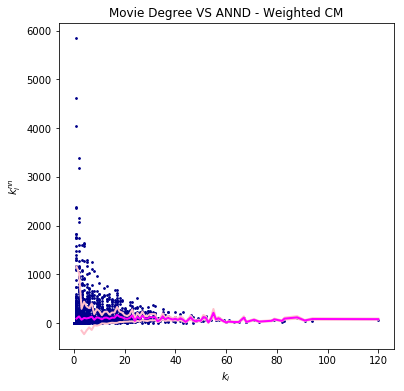}
\includegraphics[width = 3.9cm, height=3.9cm]{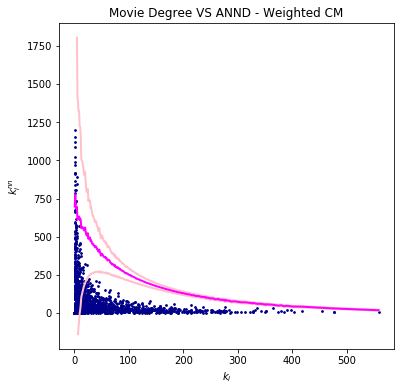}
\includegraphics[width = 3.9cm, height=3.9cm]{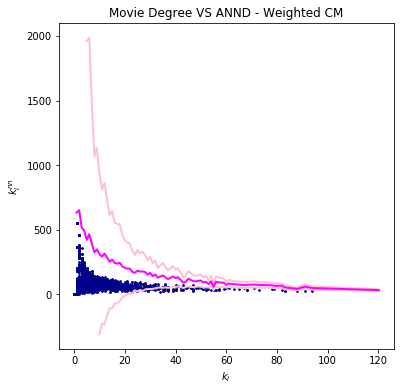}\\
\includegraphics[width = 3.9cm, height=3.9cm]{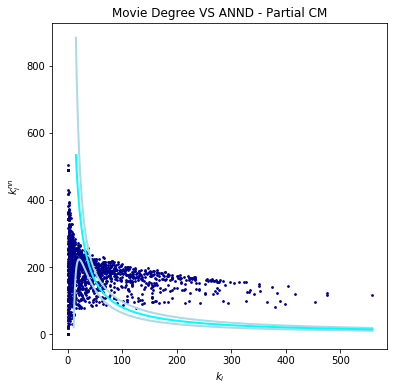}
\includegraphics[width = 3.9cm, height=3.9cm]{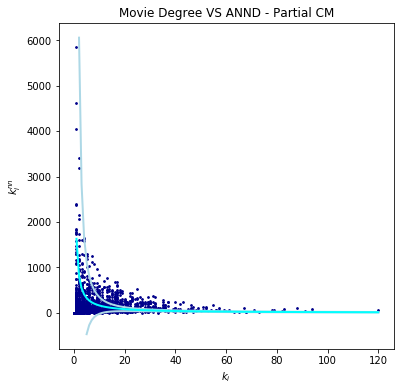}
\includegraphics[width = 3.9cm, height=3.9cm]{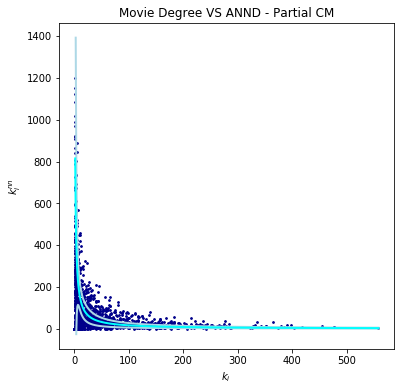}
\includegraphics[width = 3.9cm, height=3.9cm]{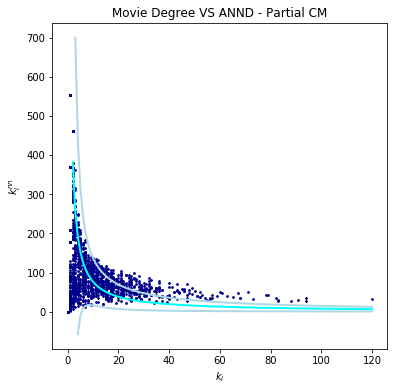}\\
\includegraphics[width = 3.9cm, height=3.9cm]{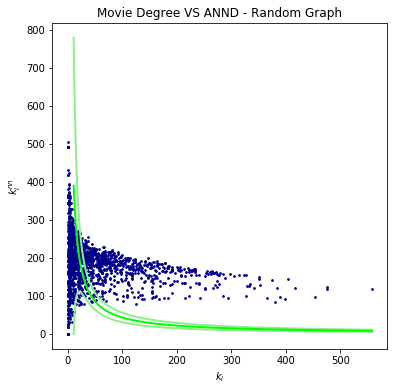}
\includegraphics[width = 3.9cm, height=3.9cm]{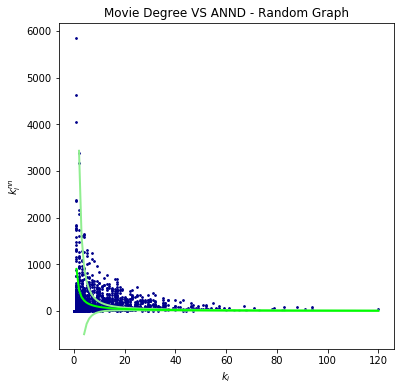}
\includegraphics[width = 3.9cm, height=3.9cm]{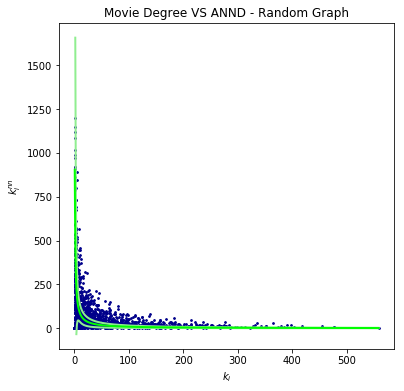}
\includegraphics[width = 3.9cm, height=3.9cm]{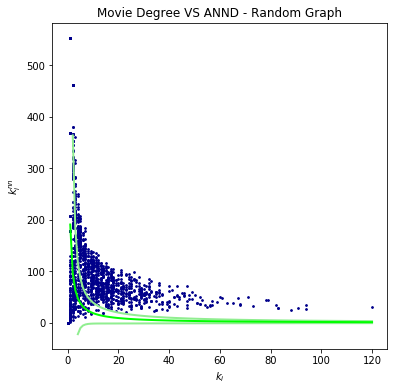}
\caption{Application of the method to the ML network. First row panels report $k_i$ versus $c_i$. In the remaining panels we have instead $k_i$ versus ANND. Left panels report $k_i^+$ versus $k_i^{pp}$. Proceeding on the right we find $k_i^-$ vs $k_i^{pn}$, $k_i^+$ vs $k_i^{np}$ and $k_i^-$ vs $k_i^{nn}$ respectively. Red lines show the expectation values computed with our method ($\pm 2$ standard deviations in orange). Magenta, blue and green lines are instead the expectations under WCM, PCM and Erd\"os-R\'enyi RG, with relative standard deviations in pink, lightblue and lightgreen.}
\label{fig:movielens - ANND}
\end{figure*}

For each dataset we employ the procedure described in the Method section to construct the benchmark model. So we obtain a set of probability matrices (one for every rating level), collecting the probability to observe rating $\beta = 1, \dots, \beta_{max}$ for each pair of nodes in the network. 

Once the Lagrange multipliers $\vec{x},\,\vec{y}$ are obtained from the maximisation of the likelihood  (\ref{eq:likelihood}), the expected quantities $\left\langle k_i^{pp} \right\rangle$, $\left\langle k_i^{pn} \right\rangle$, $\left\langle k_i^{np} \right\rangle$,  $\left\langle k_i^{nn} \right\rangle$ and $\left\langle c_i \right\rangle$ across the ensemble can be analytically computed replacing the terms $m_{i,\alpha}^+$ and $m_{i,\alpha}^-$ in equations (\ref{eq:annd_pp}-\ref{eq:checkerboard}) with their expected values $\left\langle m_{i,\alpha}^+\right\rangle$ and $\left\langle m_{i,\alpha}^-\right\rangle$. Following the instructions in \cite{squartini2011analytical}, we will also identify a confident region of two standard deviations around the average values. The comparison of observed and expected quantities indicates whether these higher order network properties can be directly explained by lower order topological structures, i.e. the constraints imposed on nodes' degrees, or require further investigation since they represent an indication of some correlation patterns in the observed network.

\begin{figure*}[ht!]
\center
\includegraphics[scale=0.5]{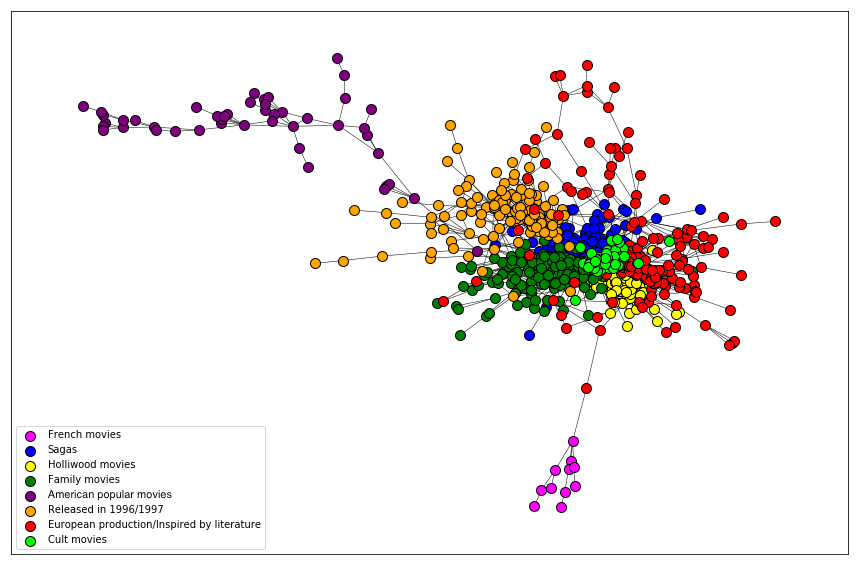}
\caption{Graphical representation of the most numerous communities for ML network. After the validation procedure, a standard modularity-based community detection algorithm is performed and the communities are here represented in different colors.}
\label{fig:movielens_communities}
\end{figure*}

Figure \ref{fig:movielens - ANND} shows the results of this comparative analysis. All the expectations have been computed averaging the values of $k_i^{pp}$, $k_i^{pn}$, $k_i^{np}$, $k_i^{nn}$ and $c_i$ over the number of nodes having the same degree. Red lines show the average connectivity and number of checkerboards estimated by our null model. Magenta, blue and green lines represent instead the same quantities estimated by WCM, Partial Configuration Model (PCM) and Erd\"os-R\'enyi Random Graph (RG), respectively. Further details about the specification of these alternative null models is provided in the Appendix. 

The overall data trend is well captured by our ensemble. However, some observations still lie outside the two standard deviations range. This may suggest the presence of extra correlations that cannot be directly traced back to the degree sequence alone, despite the full specification of scores' distribution. The analysis of the other null models would lead to completely unreliable conclusions, since in most cases, the induced ANND baseline is not able to capture the data overall trend, overestimating (especially for WCM) or underestimating (PCM and RG) the expected ANND and number of motifs in the network.

Due to the evident difference on the percentage of positive and negative observed reviews, an additional
different type of analysis has been performed on the remaining datasets, taking into consideration positive reviews only: the BiSCM outperforms even in this case, as the Appendix shows.

\begin{figure*}[ht!]
\center
\includegraphics[scale=0.5]{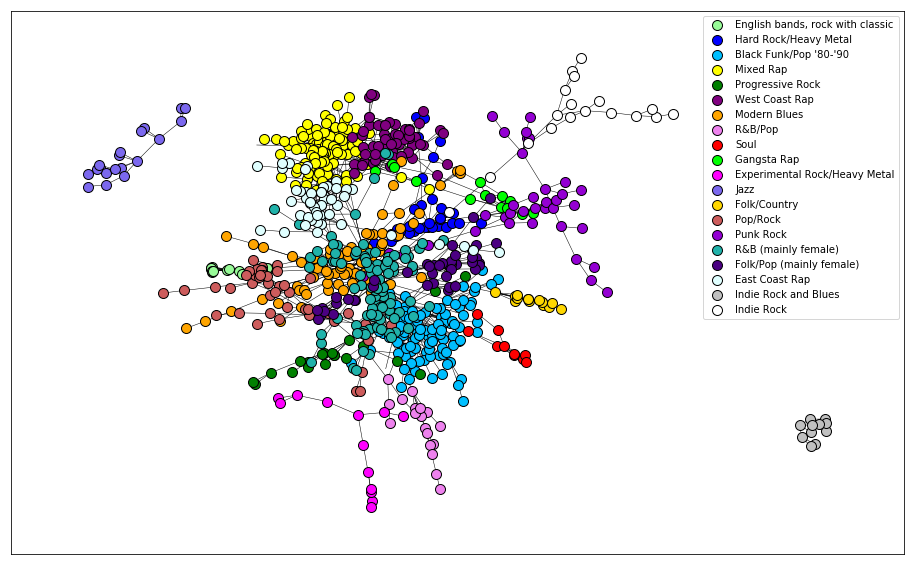}
\caption{Graphical representation of the most numerous communities for DM network. After the validation procedure, a standard modularity-based community detection algorithm is performed and the communities are here represented in different colors.}
\label{fig:music_communities}
\end{figure*}

\section{Monopartite communities}
For two of our datasets we have reported the results of another interesting performance analysis. The ML and DM networks have been binarized and then projected on the products layer, i.e. the movies and musical products layers respectively. All projection algorithms require to connect a pair of nodes in the monopartite network, whenever they share a common neighbor in the bipartite graph. However, our projected edges have been further validated using the procedure in \cite{saracco2017inferring}. The Louvain modularity-based community detection method \cite{blondel2008fast,Fortunato2010} has been finally applied to the monopartite validated projections and the final communities are here discussed.

\subsection{MovieLens}
The division in communities does not follow any genre-based division, as previously observed in \cite{saracco2017inferring}, but rather identifies some characteristics shared by the movies audience. The result of the community detection procedure is shown in Figure \ref{fig:movielens_communities}. Our method is able to detect movies released in 1996/97 the year befor the survey (in orange), such as ``Mission Impossible'', ``Independence Day'', ``Donnie Brasco''. So this group of movies is characterized by the curiosity of users towards new releases. A second group collects \emph{family movies} (as they were called in \cite{saracco2017inferring})  , including ``Cinderella'', ``101 Dalmatians'', ``Home Alone'' or ``Mrs Doubtfire'' (in green). 
In the blue community we find more ``adult'' movies, as the ``Alien" saga, the episodes of ``Die Hard", ``Escape from New York",  ``Judge Dredd", "Conan the Barbarian", as well as ``Terminator" episodes, and some westerns like ``The Good, the Bad and the Ugly" and ``Young guns".  In this community we can find even horror titles, like ``Tales From the Crypt" episodes, ``Interview with the Vampire", ``A Nightmare on Elm street" and ``Bram Stoker's Dracula". In the lime block we have cult movies, such as ``Blade Runner'', ``Star Wars'', ``Back to the Future''. Interestingly, this group collects all the available Kubrik's production (``Full Metal Jacket'', ``2001 Space Odyssey'', ``A Clockwork Orange''). The red community groups together Europen production movies (``Cinema Paradiso'', ``Mediterraneo'', ``Trainspotting''), but also movies inspired by books or theatrical plays (``Emma'', ``Hamlet'', ``Sense and Sensibility'', ``Othello''). In the last relevant block we can find classical Hollywood movies (in yellow) such as ``Casablanca'', ``Ben Hur'', ``Once upon a time in America'', ``Taxi Driver''. Interestingly, in this group we can find Hitchcock's filmography (``Vertigo'', ``Psycho'', ``Rebecca'').  Some smaller communities have however a very clear and defined characterization. For instance we have the community of all Wallace\&Gromit short animation movies or the group of French production dramas (in magenta). Our results is in substantial agreement with those of \cite{saracco2017inferring}, but for few differences: essentially, what are the cult and the horror communities in \cite{saracco2017inferring} here move to similar communities, but with a higher focus on cult movies, such that cult movies are more selected. In a similar way, the BiSCM projection is able to capture the niche of French films, while the BiCM one was not.\\ 
Such behaviour is due to the constraints imposed by the two null models. Indeed the BiSCM is more restrictive, since the degree sequence of each of the rating bipartite network is fixed, while in the BiCM the degree sequence of just positive ratings (i.e. merging the information of 3, 4 and 5 stars) is fixed, thus allowing for greater  fluctuations. This effect can be observed in the connectance of the validated projection, which is 0.87\% for BiSCM against a value of 1.17\% for the BiCM.

\subsection{\textbf{Digital Music}}
With respect to the previous ML case, we obtain smaller and more precise groups of artists. For the DM network, each community reveals a specific genre or combination of genres. A pictorial representation of the most numerous communities of the validated network is provided in Figure \ref{fig:music_communities}.

We have the small light green community with two classic rock English bands, both of them characterised by a fusion with classical arrangements (Moody Blues and Electric Light Orchestra). Different groups collect different shades of rock: the hard rock/heavy metal community is in blue (Loverboy, Alice Cooper, Van Halen, Scorpions, Deep Purple, Lynyrd Skynyrd) while the progressive rock is in green (Premiata Forneria Marconi, Soft Machine) and experimental rock in magenta. 
In dark violet we can find the grunge rock and related tendencies: Alice in Chains, Pearl Jam, Soundgarden, as well as Red Hot Chilli Peppers, Iggy and the Stooges and the MC5. In indian red there is a community including Elton John, Billy Joel, the Genesis as well as Phil Collins and Peter Gabriel in their solo career. The sea green and indigo groups represent respectively female R\&B singers (Whitney Houston, Aretha Franklin, Alicia Keys, Nelly Furtado) and female folk/pop ones (Alanis Morisette, Anastacia, Vanessa Carlton, Dido). The rap genre is divided between east coast and west coast hip hop, gangsta rap and a mixed community with the most famous artists (Eminem, Jay Z, 2 Pac, 50 Cent, D12), respectively depicted in light blue, dark magenta, lime and yellow. 
The isolated community in violet collects respectively jazz (Thelonious Monk, Miles Davis, Cannonball Adderley, Charles Mingus, Sonny Rollins), while the red one contains almost exclusively James Brown albums. A folk/country community (almost exclusively composed by John Denver and Gordon Lightfoot albums) is represented in gold. We finally have the grey and white groups with indie rock artists (Radiohead, Bon Iver, Of Monsters and Men), the R\&B singers and songwriters in pink (Marvin Gaye, Johnny Gill, Luther Vandross). In orange we find the community of folk/rock/blues, including Bob Dylan, Jimi Hendrix, Eric Clapton, the Who, Paul Simon but also the subsequent Elvis Costello, Bruce Springsteen. It is interesting to find here even Robert Johnson, the legendary bluesman, who was a source of inspiration for the artists in this community. The community of soul-funk (the Jackson 5, Barry White, Stevie Wonder, the Commodores, the Parliament, Sly and the Family Stone, Prince, the Isley Brothers) is in cyan. It is interesting to note that some Jamiroquai albums can be found in this latter community: indeed several experts compared the first production of this artist to Stevie Wonder~\footnote{\url{https://www.allmusic.com/artist/jamiroquai-mn0000176358/biography}}. Some smaller communities have not been included in the plot due to their low number of participants. However their interpretation is still clear, since they generally collect single artists (Leonard Cohen) or identify a very specific music genre (such as the group of white rappers Insane Clown Posse and Anybody Killa of genre horrorcore).

\section{Conclusions}
In everyday web experience, it is possible to find lots of different examples of online review platforms: from Amazon customer ratings, to Tripadvisor, Anobii, just to mention the most famous ones. All of these services provide an incredible source of information: indeed they are currently used for recommendation systems in order to focus possible advertisements on items close to the customer tastes~\cite{Herlocker2000,Karypis2001,Rennie2005,Salakhutdinov2007,Koren2009,Tang2016}. Nevertheless, a proper randomisation of this kind of system cannot be found in the literature, at the best of our knowledge: indeed, this kind of systems can be represented by a bipartite weighted network in which we have just a limited number of weights, i.e. the possible ratings of a review. In principle it could be tempting to randomise it using the usual standard approach used for ``unlimited" weighted networks, but the outcome could be a nonsense, like having reviews with a score higher than the maximum value. We fill this gap by considering the randomisation of finite weights network.\\ 
In order to provide an unbiased framework, we follow the research line of entropy-based null model~\cite{park2004statistical,garlaschelli2008maximum,squartini2011analytical}. In the case of score network, the difficulty resides in considering mutually exclusive outputs for each entry, i.e. having different possible scores with different probabilities. Our approach resembles the one of~\cite{mastrandrea2014enhanced} for the reciprocal configuration model. Even in that case there were 4 mutually exclusive possibilities, for every link, for every node: no link, an exclusively outgoing link, an exclusively ingoing link or a reciprocal link. Following this track, we were able to extend the Configuration Model to rating networks; indeed, as shown, our method can be applied even to signed networks, as well as for the randomisation of categorical bipartite network (in the latter case a projection on one of the layer is hard to define; we leave this study for further research).\\
The application of such a framework is able to capture some non trivial information like the abundance of topological pattern as the extensions of the $k^{nn}$ and bipartite motifs to rating networks. After observing the ability of the model of capturing the information contained in the original network, we used it in order to filter the information contained in one of the two layers: analysing the Amazon Digital Music dataset \cite{Note2}, we were able to uncover communities of music based on customers taste. Analysing the dataset of \cite{Note1} it was possible to refine the community detection of~\cite{saracco2017inferring}: indeed our model is more constrained than the one proposed there and it filters more the just the BiCM, after the binarization.\\
Our present work is just the first step of different possible applications: such a model can in fact be used in order to provide a more detailed recommendation system, through the added value of the statistical significance of the observations.

\section{Acknowledgements}
This work was supported by the EU projects CoeGSS (Grant No. 676547), MULTIPLEX (Grant No. 317532), Openmaker (Grant No. 687941), SoBigData (Grant No. 654024), and the FET projects SIMPOL (Grant No. 610704), DOLFINS (Grant No. 640772). The authors are grateful to Giulio Cimini and Tiziano Squartini for useful discussions.

\section{Appendix}
\subsection{Entropy maximisation in BiSCM}
Let us maximise the entropy (\ref{eq:entropy}) under the constraints on the degrees for each possible rating. The Hamiltonian reads as
\begin{equation*}
\begin{split}
H=&\sum_\beta \Big(\sum_i k_{i,\beta}\cdot\eta_{i,\beta}+\sum_\alpha k_{\alpha,\beta}\cdot\theta_{\alpha,\beta}\Big)\\
=&\sum_\beta \sum_{i,\alpha}m_{i,\alpha,\beta}\Big( \eta_{i,\beta}+\theta_{\alpha,\beta}\Big),
\end{split}
\end{equation*}
meaning that the expectation value of the degree per rating is conserved. As in this class of systems, the solution is quite straightforward: 
\begin{equation}\label{eq:P}
P(\textbf{G}|\vec{\eta}_\beta,\vec{\theta}_\beta)=\dfrac{e^{-H(\mathbf{G})}}{Z},
\end{equation}
where $Z$ is the partition function. The computation of the partition function returns instead
\begin{equation}\label{eq:Z}
\begin{split}
Z=&\sum_{\mathbf{G}\in\mathcal{G}}\prod_\beta\prod_{i,\alpha}(x_{i,\beta}\,y_{\alpha,\beta})^{m_{i,\alpha,\beta}(\mathbf{G})}\\
=&\prod_\beta\prod_{i,\alpha}\sum_{\mathbf{G}\in\mathcal{G}}(x_{i,\beta}\,y_{\alpha,\beta})^{m_{i,\alpha,\beta}(\mathbf{G})}\\
=&\prod_{i,\alpha}\Big(1+ \sum_\beta x_{i,\beta}\,y_{\alpha,\beta}\Big),
\end{split}
\end{equation}
where $x_{i,\beta}=e^{-\eta_{i,\beta}}$, $y_{\alpha,\beta}=e^{-\theta_{\alpha,\beta}}$ and the last step is justified by the fact that all $m_{i,\alpha,\beta}$ are mutually exclusive, so the presence of an edge with rating $\hat{\beta}$ excludes all the others (i.e. $m_{i,\alpha,\hat{\beta}}=1 \Rightarrow m_{i,\alpha,\beta}=0, \,\forall\beta\neq\hat{\beta}$). Implementing equations (\ref{eq:Z}) in (\ref{eq:P}) we get (\ref{eq:P_M}).

\subsection{Entropy maximisation in truncated WCM}
In this case the entropy (\ref{eq:entropy}) is maximised under the constraints on the observed strengths. The Hamiltonian of the problem is
\begin{equation*}
\begin{split}
H &= \sum_i s_i \cdot \eta_i + \sum_\alpha s_\alpha \cdot \theta_\alpha \\
&= \sum_{i,\alpha} m_{i,\alpha} \left(\eta_i + \theta_\alpha \right),
\end{split}
\end{equation*}
meaning that we preserve the expectation value of nodes' strenghts. Again, the solution is straightforward and given by (\ref{eq:P}). However, taking into account that $m_{i,\alpha}$ can only vary into the range $1, \dots, \beta_{max}$, the partition function $Z$ returns
\begin{equation}\label{eq:z_wcm}
\begin{split}
Z &= \sum_{\mathbf{G} \in \mathcal{G}}\prod_{i,\alpha}\left(x_i \, y_\alpha\right)^{m_{i,\alpha}(\mathbf{G})} \\
&= \prod_{i,\alpha}\sum_{\mathbf{G} \in \mathcal{G}} \left(x_i \, y_\alpha\right)^{m_{i,\alpha}(\mathbf{G})} \\
&= \prod_{i,\alpha}\frac{1-(x_i \, y_\alpha)^{\beta_{max}}}{1-x_i \, y_\alpha}, 
\end{split}
\end{equation}
where $x_i = e^{-\eta_i}$ and $y_\alpha = e^{-\theta_\alpha}$. As in the previous case, the last passage is justified by the fact that edges are mutually exclusive and the presence of a score $\hat{\beta}$ excludes all the others. With the implementation of equations (\ref{eq:P}) and (\ref{eq:z_wcm}) it is possible to get the following probability distribution on the graphs ensemble
\begin{equation*}
P(\mathbf{M}|\vec{x}, \vec{y}) = \prod_{i,\alpha}\frac{(1-x_i \, y_\alpha)(x_i \, y_\alpha)^{m_{i,\alpha}}}{1-(x_i\, y_\alpha)^{\beta_{max}+1}}.
\end{equation*}
while the probability to observe a link with rating $\beta$ between nodes $i$ and $\alpha$ reads as follows
\begin{equation}\label{eq:truncated_geom}
p_{i,\alpha, \beta} = \frac{(1-x_i\,y_\alpha)(x_i\,y_\alpha)^\beta}{1-(x_i\,y_\alpha)^{\beta_{max}+1}}.
\end{equation}
Note that the previous equation (\ref{eq:truncated_geom}) identifies a truncated geometric distribution with parameter $1-x_i y_\alpha$ and $\beta \in \{1, \dots, \beta_{max}\}$.

\subsection{Entropy maximisation in PCM}
The entropy maximisation procedure in the PCM framework follows exactly the same steps presented in the BiSCM section, but with a reduced number of imposed constraints. Indeed, in this framework we just preserve the expectation value of the degrees per rating on one single pre-defined layer. Then, the Hamiltonian reads as follows
\begin{equation}\label{eq:H_pcm}
H=\sum_{i,\beta} k_{i,\beta}\cdot\eta_{i,\beta}=\sum_\beta \sum_{i,\alpha}m_{i,\alpha,\beta}\, \eta_{i,\beta}
\end{equation}
while the partition function $Z$ becomes
\begin{equation}\label{eq:z_pcm}
\begin{split}
Z &= \sum_{\mathbf{G} \in \mathcal{G}} \prod_{\beta}\prod_{i,\alpha}(x_{i,\beta})^{m_{i,\alpha,\beta}(\mathbf{G})} \\
&= \prod_\beta \prod_{i,\alpha} \sum_{\mathbf{G} \in \mathcal{G}} (x_{i,\beta})^{m_{i,\alpha,\beta}(\mathbf{G})} \\
&= \prod_{i,\alpha} \Big(1 + \sum_{\beta} x_{i,\beta}\Big)
\end{split}
\end{equation}
where $x_{i,\beta} = e^{-\eta_{i,\beta}}$. Combining the two equations (\ref{eq:H_pcm}) and (\ref{eq:z_pcm}) we obtain the following probability distribution over the graphs ensemble
\begin{equation*}
\begin{split}
P(\mathbf{M}|\vec{x}) &= \prod_{i,\alpha} \frac{\prod_\beta (x_{i,\beta})^{m_{i,\alpha,\beta}}}{1 + \sum_\beta x_{i,\beta}} \\
&= \prod_i \frac{\prod_\beta (x_{i,\beta})^{k_{i,\beta}}}{(1 + \sum_{\beta}x_{i,\beta})^{N_\Lambda}}
\end{split}
\end{equation*}
where the single term inside the product 
\begin{equation}
p_{i,\beta} = \frac{x_{i,\beta}}{1 + \sum_\beta x_{i,\beta}} = \frac{k_{i,\beta}}{N_\Lambda}
\end{equation}
simply identifies the probability to observe a link with score $\beta$ incident to node $i$ and coincides with the empirical observed frequency for score $\beta$. Notice that the previous model has been defined considering the degrees $k_{i,\beta}$ for $i \in \{1, \dots, N_L\}$ as constraints. However, the analogous counterpart can be implemented imposing the degrees of nodes on the other layer $k_{\alpha}$, for $\alpha \in \{1, \dots, N_\Lambda\}$.

\subsection{Entropy maximisation in Erd\"os-R\'enyi Random Graph}
In the last considered null model, the entropy is maximised under the constraint on the observed number of edges per score $E_\beta$ only. The Hamiltonian of the problem is
\begin{equation}\label{eq:h_rg}
H = \sum_\beta \theta_\beta \cdot E_\beta(\mathbf{G}) = \sum_\beta \theta_\beta \sum_{i,\alpha} m_{i,\alpha\beta}
\end{equation}
meaning that, for each score, we want to preserve the observed number of edges, while the partition function $Z$ reads as follows
\begin{equation}\label{eq:z_rg}
\begin{split}
Z &= \sum_{\mathbf{G}\in \mathcal{G}} \prod_{i,\alpha,\beta} x_\beta \\
&= \prod_{i,\alpha} \Big(1 + \sum_{\beta} x_\beta \Big)\\
&= \Big(1 + \sum_{\beta}x_\beta \Big)^{N_L N_\Lambda}
\end{split}
\end{equation}
where $x_\beta = e^{-{\theta_\beta}}$. Implementing equations (\ref{eq:h_rg}) and (\ref{eq:z_rg}) into equation (\ref{eq:P}) we obtain the following probability distribution
\begin{equation*}
P(\mathbf{M}|\vec{x}) = \frac{\prod_\beta x_\beta^{E_\beta}}{\Big(1 + \sum_\beta x_\beta \Big)^{N_L N_\Lambda}}
\end{equation*}
where the term
\begin{equation}
p_\beta = \frac{x_\beta}{1 + \sum_\beta x_\beta} = \frac{E_\beta}{N_L N_\Lambda}
\end{equation}
denotes the probability to observe a link of score $\beta$ between any pair of nodes. Notice that the previous probability is invariant for all pairs of nodes and coincides with the empirical frequency of observed edges for the considered rating.

\subsection{Chung-Lu Approximation}

\begin{figure*}[th!]
\center
\includegraphics[scale=0.3]{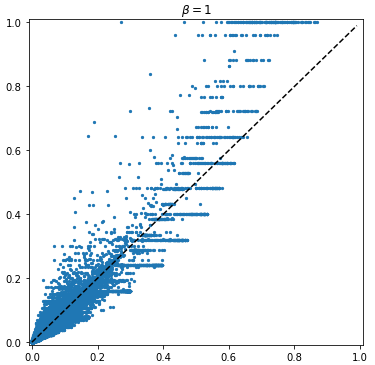}
\includegraphics[scale=0.3]{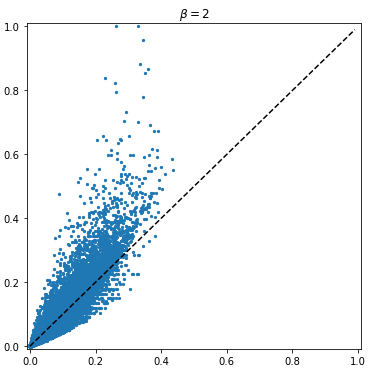}
\includegraphics[scale=0.3]{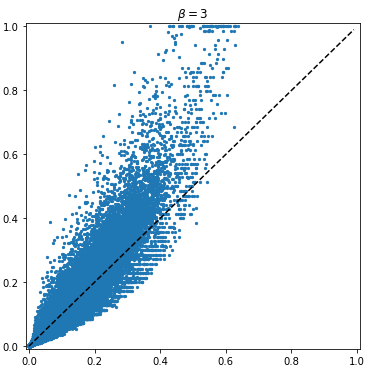}\\
\includegraphics[scale=0.3]{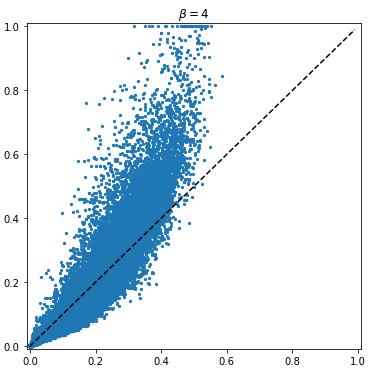}
\includegraphics[scale=0.3]{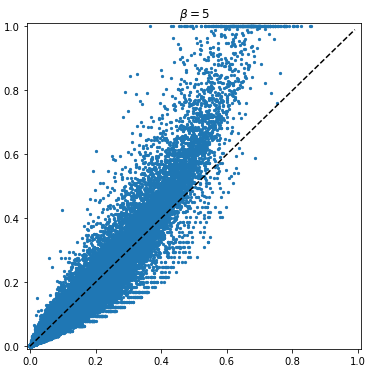}
\caption{Graphical comparison of the two definitions of probabilities in the ML network. The plot shows $p_{i,\alpha,\beta}$ on the $x$-axis and $p_{i,\alpha,\beta}^{CL}$ on the $y$-axis.}
\label{fig:chung_vs_me}
\end{figure*}

To relax the constraints required \cite{chung2002connected}, we define the connection probability between each pair of nodes in the network as
\begin{equation}\label{eq:chung_lu_probs}
p_{i,\alpha,\beta}^{CL} = \frac{k_{i,\beta}\,k_{\alpha,\beta}}{M_\beta}
\end{equation}
for all $i = 1, \dots, N_L$, $\alpha = 1, \dots, N_\Lambda$ and $\beta = 1, \dots, \beta_{max}$. The term $M_\beta = \sum_{i,\alpha} m_{i,\alpha,\beta}$ in equation (\ref{eq:chung_lu_probs}), identifies the total number of observed links for each score present in the data. Figure \ref{fig:chung_vs_me} provides a graphical comparison of the two definitions of probability for the ML network. It is evident that equation (\ref{eq:chung_lu_probs}) systematically overestimates the BiSCM values, especially for high probabilities. 

\bibliographystyle{apsrev4-1}
%

\end{document}